\documentclass[twocolumn,aps,floatfix]{revtex4}
\usepackage{amssymb}
\usepackage{graphicx}
 
\begin{document}
 
\title{Temperature dependent Bogoliubov approximation in the classical 
       fields approach to weakly interacting Bose gas}
 
\author{Miros{\l}aw Brewczyk,$\,^1$ Peter Borowski,$\,^2$
        Mariusz Gajda,$\,^3$ and Kazimierz Rz{\c a}\.zewski$\,^2$}
 
\affiliation{
\mbox{$^1$ Uniwersytet w Bia{\l}ymstoku, ulica Lipowa 41,
                        15-424 Bia{\l}ystok, Poland}  \\
\mbox{$^2$ Centrum Fizyki Teoretycznej PAN, Aleja Lotnik\'ow 32/46, 
                        02-668 Warsaw, Poland} \\  
\mbox{$^3$ Instytut Fizyki PAN, Aleja Lotnik\'ow 32/46,
                        02-668 Warsaw, Poland} }
\date{\today} 

\begin{abstract}

A classical fields approximation to the finite temperature microcanonical 
thermodynamics of weakly interacting Bose gas is applied to the idealized 
case of atoms confined in a box with periodic boundary conditions. We 
analyze in some detail the microcanonical temperature in the model. We 
also analyze the spectral properties of classical amplitudes of the plane 
waves -- the eigenmodes of the time averaged one--particle density matrix. 
Looking at the zero momentum component -- the order parameter of the 
condensate, we obtain the nonperturbative results for the chemical 
potential. Analogous analysis of the other modes yields nonperturbative 
temperature dependent Bogoliubov frequencies and their damping rates. 
Damping rates are linear functions of momenta in the phonon range and
show more complex behavior for the particle sector. Where available, 
we make comparison with the analytic estimates of these quantities. 

\end{abstract}

\maketitle

\section{Introduction}
Since the achievement of Bose-Einstein condensation in dilute atomic gases \cite{BEC}
there is a remarkable experimental activity in this area of physics. Quantum properties 
of the weakly interacting Bose gas are studied both close and away from thermal 
equilibrium. While in most cases the experiments are performed with as cold gas sample 
as possible, some experiments test the dependence of these properties on temperature. 
The central role in the theoretical studies of this system is played by the collective 
excitations. At and near zero temperature these collective excitations are described 
very well by the Bogoliubov approximation. Much less is known about the collective 
excitations at higher temperatures.

More generally, quantum theory of weakly interacting Bose gas at finite temperatures is
still a challenge. To answer this challenge
most researchers describe the system as consisting of two different,
mutually interacting components: the condensate and the thermal cloud \cite{Stringari}. 
This way a
significant progress has been achieved in explaining such experimental observations as,
for instance, temperature shifts and damping rates of various oscillation modes of the 
two component system \cite{Zaremba}. The two component approaches, although successful, 
are unsatisfying in their underlying assumptions. The splitting of the gas at nonzero 
temperature into the condensate and the thermal cloud should be the result of the Bose 
statistics and of interactions. 

In a series of papers several groups formulated the, so called, classical fields
approximation \cite{classf,OpEx,Davis,Oxford1,Oxford2,RapCom,Castin}. 
Within this formulation there are, however, two slightly different
approaches. The group of ENS \cite{Castin} developed the so called
truncated Wigner method for Bose condensed gases. The main idea is to
describe the total system by a classical field obeying the Gross--Pitaevskii 
equation. The classical field, except of a condensate
component, contains also a stochastic part representing a thermal
cloud. Mean values of observables are calculated as averages over an
ensemble of classical fields which are chosen according to the Wigner
quasi--distribution function of the initial thermal equilibrium density
operator of the gas. This approach corresponds to the canonical
description of the system where a temperature not energy is a control
parameter.

The approach that we use has been formulated in \cite{OpEx,Oxford1} and
corresponds to the microcanonical description.  This approximation,
drawing from quantum optics, consists of replacing the quantum
mechanical operators of highly occupied modes of the Bose field with
complex c--number amplitudes. The idea of replacing the annihilation
(creation) operators by c--number amplitudes is a straightforward
generalization of Bogoliubov hypothesis to a case when large number of
modes is macroscopically occupied.
It is the aim of this paper to analyze the classical
fields approximation in terms of such fundamental notions as the microcanonical
temperature, the chemical potential and the quasiparticle excitations. 

The paper is 
organized as follows: In Section \ref{class}, for the sake of completeness, we briefly
describe the classical fields approximation. In Section \ref{approx} we analyze the 
dynamical equations of the approximation from the point of view of underlying spectra. 
In Section \ref{num} the numerical results are discussed to support the analysis of 
Section \ref{approx}. In particular we show that the Bogoliubov energies at high 
temperature follow the gapless formula derived in \cite{Pethick}. We also analyze
the thermal damping rates of the modes. We find that they are proportional to 
momenta for the phonon excitations. The dependence of damping rates of particle--like 
modes on momentum is more complex showing the existence of bending points.
We present some concluding remarks in Section \ref{concl}.

\section{Classical fields approximation}
\label{class}
We consider $N$ identical atoms subject to Bose-Einstein statistics, confined to a 
cubic box of volume $V$ , with their atomic wave functions satisfying periodic boundary 
conditions. The atoms interact via contact potential characterized by the scattering 
length $a_s$. We use the method of second quantization. Our dynamical variable is a 
bosonic field operator $\hat{\Psi}({\bf r},t)$ that destroys a particle at position 
${\bf r}$ and obeys standard bosonic commutation $[\hat{\Psi}({\bf r},t),
\hat{\Psi}^{\dagger}({\bf r}',t)] = \delta({\bf r}-{\bf r}')$ relations.  
The Hamiltonian of the system reads:
\begin{eqnarray}
H &=& \int d^{\,3}r \; \hat{\Psi}^{\dagger}({\bf r},t) \frac{p^2}{2m}
\hat{\Psi}({\bf r},t)    \nonumber \\
&+& \frac{2 \pi \hbar^2 a_s}{m}
\int d^{\,3}r \; \hat{\Psi}^{\dagger}({\bf r},t) \hat{\Psi}^{\dagger}({\bf r},t)
\hat{\Psi}({\bf r},t) \hat{\Psi}({\bf r},t)  \;.   \nonumber \\
\label{Hamil}
\end{eqnarray}

The Heisenberg equation of motion for the field operator follows:
\begin{eqnarray}
i \hbar \frac{\partial \hat{\Psi}({\bf r},t)}{\partial t} =
&-& \frac{\hbar^2}{2m} \nabla^2 \, \hat{\Psi}({\bf r},t)   \nonumber \\
&+&
\frac{4 \pi \hbar^2 a_s}{m} \; \hat{\Psi}^{\dagger}({\bf r},t) \hat{\Psi}({\bf r},t)
\hat{\Psi}({\bf r},t)   \;.
\label{Heis}
\end{eqnarray}
We do not know how to solve this nonlinear operator equation. However, a natural
simplification is possible for the degrees of freedom (modes) of the field, which are
occupied by large number of atoms. Just like corresponding highly occupied modes of
electromagnetic field, they can be described by c--number complex amplitudes rather than
by their creation and annihilation operators. On the other hand the modes that are
sparsely populated and require full quantum treatment, may be in the crudest
approximation neglected.

The symmetry of the box with periodic boundary condition determines a natural set of
modes for the thermal equilibrium states of atoms. The natural modes are simply the
plane waves, with quantized momentum ${\bf p} = 2\pi \hbar (n_1,n_2,n_3) /L$
($n_i$ being integer). Therefore, the field operator can be expanded in these modes:
\begin{equation}
\hat{\Psi}({\bf r},t) = \frac{1}{\sqrt{V}} \sum_{{\bf p}} \exp(-i {\bf p} {\bf r} /\hbar) 
\hat{a}_{{\bf p}}(t)  \;.
\label{expan}                                
\end{equation}
Substituting the expansion (\ref{expan}) into (\ref{Heis}) we get a set of nonlinear 
equations for the creation and annihilation operators of the plane wave modes:
\begin{equation}
\frac{d \hat{a}_{{\bf p}}(t)}{dt} = -i \frac{p^2}{2m \hbar} \hat{a}_{{\bf p}}
-i \frac{g}{V\hbar} \sum_{{\bf q}_1, {\bf q}_2} 
\hat{a}^{\dagger}_{{\bf q}_1} \hat{a}^{}_{{\bf q}_2}
\hat{a}^{}_{{\bf p}+{\bf q}_1-{\bf q}_2}  \;,
\label{eqnoper}                                
\end{equation}
where the coupling $g$ is equal to:
\begin{equation}
g = \frac{4\pi \hbar^2 a_s}{m}  \;.
\label{g}                                                                                  
\end{equation}

At this point we identify the set of highly occupied modes. By definition these are all
long wave length degrees of freedom up to a maximal cut--off momentum ${\bf p}_{max}$.
For these modes we replace the operators with complex amplitudes:
\begin{equation}
\hat{a}_{\bf p} = \sqrt{N} \alpha_{\bf p}
\label{ampl}                                                                                  
\end{equation}
in such a way that the square of the modulus of the amplitude gives the fraction of 
atoms in a given mode. 
The above replacement is a direct generalization of the Bogoliubov hypothesis.
   
Confining our attention to the highly occupied modes we are 
reducing the dynamical equations to the set of nonlinear differential equations for 
the amplitudes:
\begin{equation}
\frac{d \alpha_{{\bf p}}(t)}{dt} = -i \frac{p^2}{2m \hbar} \alpha_{{\bf p}}
-i \frac{g N}{V\hbar} \sum_{{\bf q}_1, {\bf q}_2} 
\alpha^{*}_{{\bf q}_1} \alpha^{}_{{\bf q}_2}
\alpha^{}_{{\bf p}+{\bf q}_1-{\bf q}_2}  \;,
\label{eqnalpha}                                
\end{equation}
which may be solved numerically. 
Note that summation over momenta is 
truncated at $|{\bf q_1}|= |{\bf q_2}|=p_{max}$, where $p_{max}$ is
a value of momentum of the highest mode which occupation is still
macroscopic. In numerical implementation of the method the value of
$p_{max}$ has to be determined self-consistently.

Eq. (\ref{eqnalpha}) is a momentum representation of the standard
Gross--Pitaevskii equation. For numerical purposes it is convenient to
switch to position representation which corresponds to the c--number
version of the operator Eq. (\ref{Heis}) on rectangular grid with the grid
spacing defined by a cut--off momentum $\Delta x=\pi/p_{\max}$. In the classical
fields method, a finite grid version of the Gross--Pitaevskii equation
describes a total system -- Bose-Einstein condensate and a thermal
cloud.

This interpretation is shared by other groups which apply the classical
fields method for microcanonical description of a cold Bose system
\cite{Oxford1,Oxford2,Davis}. In the recent paper  \cite{Davis} the value of 
the cut--off momentum has been chosen arbitrary. It means that only a fraction
of all macroscopically occupied modes is taken into account, therefore
in such a treatment there is no simple method of  determination of a
number of particles in the whole system. This leads to difficulties in
assigning  an absolute value of temperature and authors of \cite{Davis}
calculate some scaled temperature $T/N$ rather than $T$.
We treat the problem of a cut--off momentum differently. In the present 
approach a number of classical
modes, or equivalently a cut--off momentum, is an important physical
parameter. We determine its value by requiring
that an occupation of the cut--off momentum mode is equal to one. This
is a kind of compromise as the validity of approximation is limited to
modes with occupation  large  compared to one, but on the other hand
we want to describe the whole system. Such a choice of the cut--off
momentum ensures that the total occupation of modes with momenta not
included in our computations is only a few percent even close to
critical temperature. Moreover, we are able
to determine the absolute (not scaled) value of  temperature of the
system.
        
The most important finding is \cite{OpEx,Oxford1} 
that almost any 
initial condition of given number of atoms and given energy, after the transient 
thermalization period leads to a steady state that does not depend on the particular 
choice of initial conditions. This way we are able to produce a numerical version of 
fluctuating, weakly interacting Bose gas in the quantum degenerate region. Of course, 
at temperatures close to the critical one there are many highly occupied modes of 
the Bose field. All authors using the classical fields approximation identify the 
zero momentum component of the wave function with the condensate and all other 
components as a part of the thermal cloud. In a recent paper \cite{RapCom} we have 
investigated this problem and came to the conclusion that the coarse graining due 
to the finite resolution of the realistic measurement, both in time and in space, 
reduces the pure state described by a single wave-function of the method to a
mixed state in which there is no coherence between various momentum components of
the field. Only with this interpretation the perfectly isolated system of a 
microcanonical approach may be viewed in quantum mechanics as a mixed state.
In the next Sections we are going to spectrally analyze the long time solutions of 
Eqs. (\ref{eqnalpha}) to generalize the celebrated Bogoliubov approximation to higher 
temperatures.

\section{Excitation spectrum: approximate treatment}
\label{approx}
In this section we present an approximate analytical treatment of 
dynamical equations of the classical fields method. Under assumption of
a steady--state evolution we discover some constants of motion what 
allows us to linearize the equations and find an excitation spectrum 
of the system.  This approach leads to the notion of quasiparticles 
and the famous Bogoliubov--like spectrum.

The equations for complex amplitudes originating from the many body
Hamiltonian are given by (\ref{eqnalpha}) and in a particular units
of length (the size of the box $L$) and time ($mL^2/\hbar$)
are:
\begin{equation}
\dot{\alpha}_{\bf p} = -i {{\bf p}^2 \over 2} \alpha_{\bf p}
-igN \sum_{{\bf l}, {\bf m}} \alpha^*_{\bf l} \alpha^{}_{\bf m} 
\alpha^{}_{\bf p+l-m} \;.
\label{dyn_eq}
\end{equation}
The set of nonlinear Eqs. (\ref{dyn_eq}) has two constants of motion. 
The first is the total energy of the system,
$E=N\sum_{{\bf k}}({\bf k}^2/2) n_{{\bf k}}+(gN/2)\sum_{\bf k,l,m}\alpha^*_{\bf
k} \alpha^*_{\bf l} \alpha^{}_{\bf m} \alpha^{}_{\bf k+l-m}$ and the second is the 
number of particles $N \sum_{\bf k} n_{\bf k}$, where 
$n_{\bf k}=\alpha^*_{\bf k} \alpha^{}_{\bf k}$.
As we have already mentioned, the system reaches a thermal
equilibrium \cite{OpEx}
after  some transient time, which depends on constants of motion only. In this 
state, a condensate
occupation undergoes small fluctuations around a well defined mean
value. Neglecting these small fluctuations we assume that $n_0 =
\alpha^{*}_0 \alpha^{}_0$ is constant. A value of $n_0$ depends on the energy 
of the system and is close to one for small energies and decreases with
increasing energy.  

By rearranging terms in Eq. (\ref{dyn_eq}) we obtain the following
equation for the amplitude of the condensate (${\bf p}=0$) mode:
\begin{equation}
\dot{\alpha}_{\bf 0}  =
-igN(2-n_0)\alpha_0 -igN\left(\sum_{{\bf k \neq 0}}\alpha_{\bf k} \alpha_{{\bf
-k}}\right)\alpha_0^*+f_{\bf 0}(t) \;,
\label{a0}
\end{equation}
where 
\begin{equation}
f_{\bf 0}(t)=-igN \sum_{\bf k \neq l; k,l \neq 0} \alpha^*_{\bf k} \alpha_{\bf l}
\alpha_{\bf k - l}
\label{f0}
\end{equation}
plays a role of an external force. Eqs. (\ref{a0}) and its complex
conjugate describe a pair of coupled  harmonic oscillators driven by
external forces. The coupling term, $\sum_{{\bf k \neq 0}}\alpha_{\bf k} 
\alpha_{{\bf -k}}$, is an anomalous density.

\begin{figure}[hbt]
\resizebox{2.8in}{2.4in}
{\includegraphics{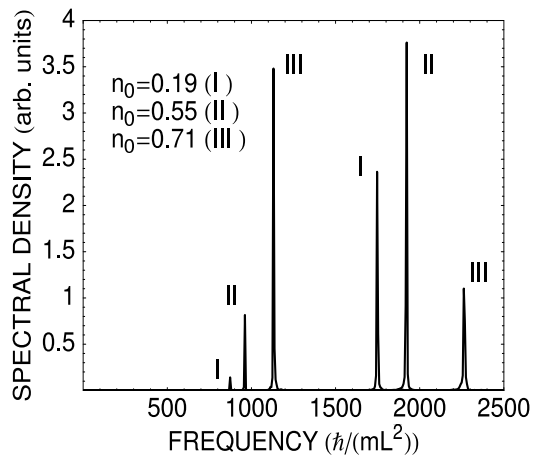}}
\caption{Fourier transform (in arbitrary units) of the anomalous density 
(peaks in the right part of the figure) and  driving force (peaks in the
left part of the figure) for various condensate occupations as indicated 
in the legend.}
\label{ansatz}
\end{figure}

In general time dependence of both the anomalous density and driving
force can be very complicated. However, in a steady state both these
quantities simply oscillate in time:  
\begin{eqnarray}
\label{delta}
\sum_{{\bf k \neq 0}}\alpha_{\bf k} \alpha_{{\bf -k}}& = & -\delta {\rm e}^{-2i(\phi
+ \mu t)} \;,\\
\label{kappa}
f_{\bf 0} & = & - \kappa \sqrt{n_0} {\rm e}^{-i (\phi +\mu t)} \;,
\end{eqnarray}
where $\mu$ plays a role of a chemical
potential (as it can be seen later), $\phi$ is some constant phase
while $\delta$ and $\kappa$ 
are (real) constants of motion depending on the energy and number 
of particles only. To support the assumed ansatz in Fig. \ref{ansatz} we 
show the Fourier transforms of the anomalous density and the driving
force for  three different values of the total energy. Indeed,
independently of  the energy, the Fourier  transform of the anomalous 
density is sharply peaked at frequency which is larger by the factor 
of two than frequency corresponding to a peak in the spectrum of the 
driving force. Because of Eqs. (\ref{delta}) and (\ref{kappa}) the equation 
for condensate amplitude has a simple  solution:  
\begin{equation}
\alpha_{\bf 0}   =  \sqrt{n_0} {\rm e}^{-i (\phi+\mu t)} \;,
\end{equation}
what is consistent with our assumption that condensate population $n_0$ 
does not depend on time. Evidently, $\mu$ plays a role of a single particle 
energy in a condensate phase. The Eq. (\ref{a0}) gives  also relation between 
chemical potential $\mu$ and other, introduced above parameters: 
\begin{equation}
\label{mu}
\mu = gN(2-n_0) - gN(\delta+\kappa)  \;.
\end{equation}
The anomalous density $\delta$ is small at low energies or
close to a critical energy, having maximum in between. The parameter
$\kappa$ is very small at low energies but it grows continuously and
gives a significant correction to the chemical potential at large
energies. Our result can be compared to the one obtained from the two 
gas model $\mu=gN(2-n_0)$ (see Fig. \ref{chempot}). Only at very small 
energies this model predicts a value of chemical potential
correctly. 

In Fig. \ref{chempot} we show the chemical potential as a
function of a condensate occupation (which is a monotonic function
of temperature). We present two curves, both corresponding to the
same value of the effective coupling $gN=731.5$, but for two
different grid sizes. The number of grid points is equal  to a number
of macroscopically occupied modes. If we increase the number of
macroscopically occupied modes while keeping population of
the highest mode constant, we  inevitably increase the total number of
particles in the system, and simultaneously decrease the
value of  $g$ ($gN$ is fixed) \cite{Schmidt}. Therefore, the 
curve corresponding to the larger grid (circles) describes 
larger system with weaker interactions as compared to the curve marked
by boxes. In Fig. \ref{chempot1} we present the chemical potential
as a function of the interaction strength $g$ at fixed condensate
fraction $n_0=0.55$ and fixed effective coupling $gN=731.5$. 
According to the two gas model the chemical 
potential should be constant in such a case and its value, 
for chosen parameters, equals $\mu =1060$.  This value is indicated
in Fig. {\ref{chempot1} by horizontal dashed line. Our results show that 
chemical potential depends on the interaction strength not only through
the product $gN$. We expect that two gas model result can be reached
in the limit of $g \rightarrow 0$, however calculation in this regime
is a demanding task. In the inset we show the chemical potential versus 
condensate population (temperature) for fixed value of both: particle 
number $N$ and interaction strength $g$. The difference between the two 
gas model prediction and our result grows with temperature.

\begin{figure}[hbt]
\resizebox{2.8in}{2.4in}
{\includegraphics{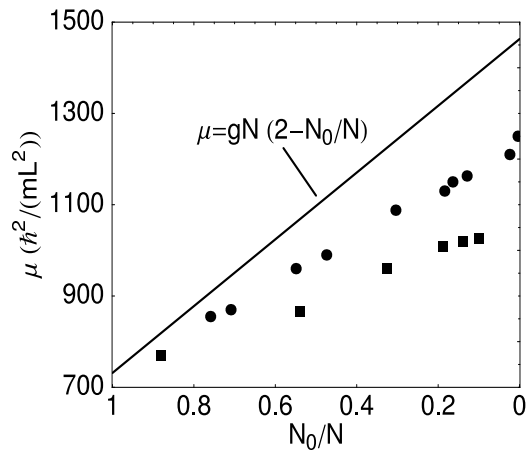}}
\caption{Chemical potential as a function of the condensate population
for the effective coupling strength $gN=731.5$. Circles and boxes show
numerical results obtained on the grids 36x36x36 and 16x16x16, respectively. 
The solid line comes from the formula derived within the two gas model. }
\label{chempot}
\end{figure}

\begin{figure}[hbt]
\resizebox{3.8in}{2.5in}
{\includegraphics{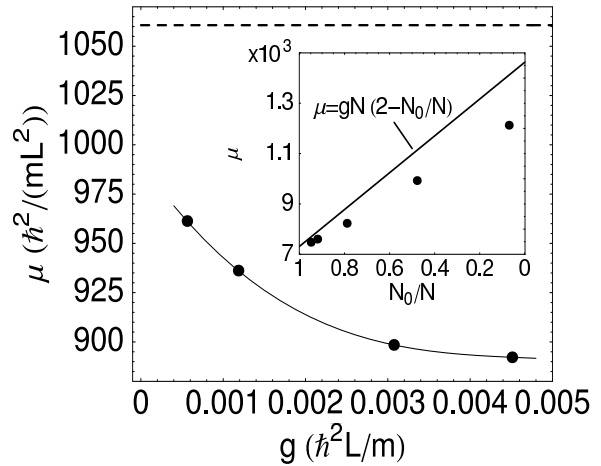}}
\caption{Chemical potential as a function of the interaction strength $g$
alone for $gN=731.5$ and the condensate occupation $n_0=0.55$ (solid line
is added to guide an eye). The horizontal dashed line indicates the value 
of the chemical potential obtained from the two gas model formula: 
$\mu = gN(2-n_0)\approx 1060$. The inset shows, just like Fig. \ref{chempot}, 
the chemical potential as a function of the condensate population but for 
a constant number of atoms $N=235000$. }
\label{chempot1}
\end{figure}

Now, we rearrange terms in the dynamical equation (\ref{dyn_eq}) for
amplitudes of excited modes to get the following expression: 
\begin{eqnarray} 
\label{ak}
\dot{\alpha}_{\bf p}=&-&i\left({{\bf p}^2 \over 2} 
+ 2gN \right)\alpha_{\bf p}    \nonumber  \\
&-&igN\left(\alpha_0^2 + \sum_{{\bf k} \neq 0}\alpha_{\bf k} \alpha_{{\bf
-k}}\right)\alpha_{{\bf -p}}^*  +f_{\bf p} \;,
\end{eqnarray}
where
\begin{equation}
f_{\bf p}=-igN \sum_{\bf k \neq -p,l; l\neq p} \alpha^*_{\bf
k} \alpha_{\bf l} \alpha_{\bf p+k - l}
\end{equation}
is a driving force of the $\alpha_{\bf p}$ mode. Guided by the
previous experience, we expect that the driving force $f_{\bf p}$ becomes
important at larger energies. However, its time dependence is not 
as simple as that of $f_{\bf 0}$. In the first approach we  omit this force in
our analytical treatment. We are well aware that doing this we loose very
important contribution to the dynamics, nevertheless even this
oversimplified  analysis gives interesting physical insight.  

After all the above simplification we  get the set
of coupled equations for amplitudes of $\alpha_{\bf p}$ and $\alpha^*_{\bf -p}$
modes. They describe excitation of the system by creation of correlated pair of
particle with momentum {\bf p} and a hole in the condensate with opposite
momentum. The solution of this set reminds famous Bogoliubov transformation: 
\begin{equation}
\label{trans}
\alpha_{\bf p}={\rm e}^{-i\mu t}\left({\rm e}^{-i\epsilon_p t}\beta_{\bf p} + 
{\rm e}^{i\epsilon_p t}\frac{\epsilon_p-\omega_p}{gN(n_0-\delta)}
\beta_{{\bf -p}}^*\right)  \;,
\end{equation}
where amplitudes $\beta_{\bf p}$ and $\beta_{\bf -p}^*$ are some
constants,  and $\epsilon_p$ is the temperature dependent Bogoliubov spectrum:
\begin{equation}
\label{spectrum}
\epsilon_p= \sqrt{\omega_p^2 - [gN(n_0-\delta)]^2}
\end{equation}
and $\omega_p$ is:
\begin{equation}
\omega_p = {p^2 \over 2}+gN(n_0+\kappa+\delta) \;.
\end{equation} 
The amplitude of the $\alpha_{\bf p}$ mode  
is a superposition of two components: the one oscillating with frequency
$\mu+\epsilon_p$ and the second with frequency $\mu-\epsilon_p$. Excitations
are Bogoliubov quasiparticles. We see, that this well known fact is
reproduced by the classical fields method and is also valid at higher
temperatures and for large interaction strength (in fact it is valid
up to a critical temperature what we will show in the next section). At low
momenta both frequency components of the amplitude $\alpha_{\bf p}$ are comparable
while at high momenta  the amplitude of the negative frequency component 
vanishes. Eq. (\ref{spectrum}) predicts a gap, i.e., the energy of
excitation does not tend to zero as the momentum ${\bf p}$ tends to
zero. This result violates the Hugenholtz-Pines  \cite{gapless}
theorem which shows 
that excitation spectrum is gapless. The reason is that in our
analysis we have neglected the term $f_{\bf p}$. Inclusion of this
term gives the correct gapless spectrum and allows for determination
of Landau damping rates of excitations.

\section{Numerical results}
\label{num}
In this section we show results obtained from numerical solution of the
dynamical Eqs. (\ref{dyn_eq}) and compare them with approximate
analytical solutions. We focus on the thermodynamics of a uniform weakly 
interacting Bose gas in the equilibrium. Knowing the spectrum of the
elementary excitations (quasiparticles) as well as their populations 
we deliver the scheme which assigns the temperature to the system. After 
that most of the thermodynamic properties of the system can be derived, 
for example, the condensate occupation as a function of temperature. On 
the other hand, the width of the quasiparticles originating from the finite 
lifetime of the elementary excitations allows one to investigate dissipative 
processes in the system.

As we have already mentioned in Section \ref{class} instead of solving 
Eqs. (\ref{dyn_eq}) it is more convenient to transform them to the position
representation what leads to the finite--grid version of the time--dependent 
Gross--Pitaevskii equation for a system of particles in a box with periodic 
boundary conditions. The equation is written as:
\begin{equation}
i \frac{\partial \Psi}{\partial t} = - \frac{1}{2} \nabla^2 \Psi
+ gN |\Psi|^2 \Psi   \;,
\label{GP}
\end{equation}
where particular units for the length (which is the box size $L$) and the time 
$(mL^2/\hbar)$ are assumed and the coupling constant $g$ equals $4 \pi a_s / L$
($a_s$ being the scattering length). This equation has to be solved on a 
rectangular grid and the grid spacing $\Delta x$ determines the value of the cut--off
momentum $p_{max}=\pi/\Delta x$.
The initial wave function is generated from the 
ground state solution by its random disturbance followed by the normalization. 
The strength of the disturbance determines the total energy per particle. Both the 
energy and the particle number are preserved during the evolution of the 
Gross--Pitaevskii equation and the system 
reaches the same stationary state independently of the choice of the initial wave 
function provided that the energy and the particle number are kept constant.

\begin{figure}[thb]
\resizebox{2.8in}{3.0in}
{\includegraphics{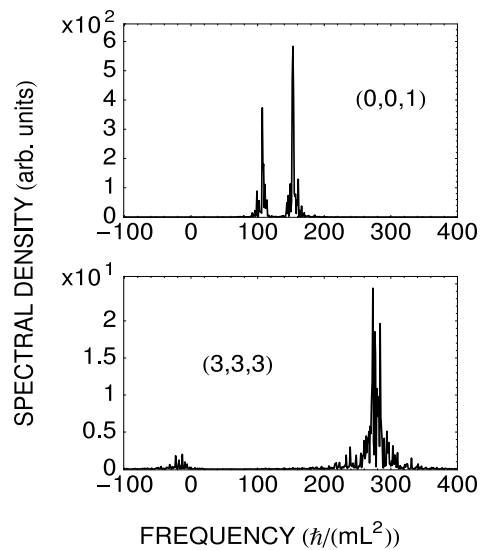}}
\caption{Fourier transform (in arbitrary units) of the (0,0,1) and
(3,3,3) modes for gN=731.5 and $n_0=0.79$ (low temperature). Note the
existence of two groups of frequencies. }
\label{spec1}
\end{figure}

\begin{figure}[hbt]
\resizebox{2.8in}{3.0in}
{\includegraphics{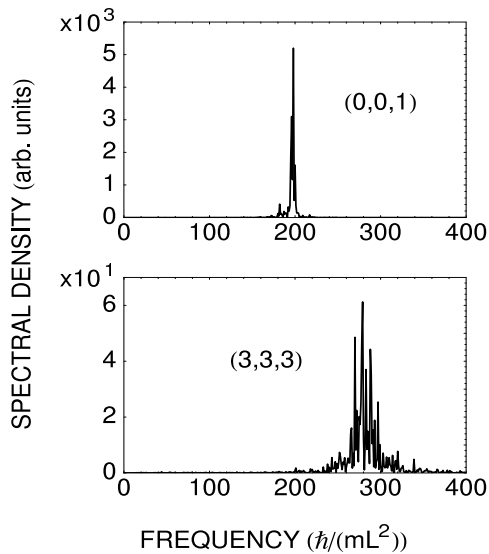}}
\caption{Fourier transform (in arbitrary units) of the (0,0,1) and
(3,3,3) modes for gN=731.5 and $n_0=0.07$ (high temperature).}
\label{spec2}
\end{figure}

As it was discussed in the previous sections, the high energy solution of
the Gross--Pitaevskii equation represents a pure state of the system and only
time averaging procedure related to the nature of the observation process
allows one to split the system to condensed and non--condensed parts. To this
end, the time averaged (after the system enters the stationary state)
single--particle density matrix is diagonalized and coherent modes (among 
them the condensate mode) and their populations are obtained. For a uniform
gas the eigenfunctions of the single--particle density matrix are just the
plane waves and the condensate is identified as the mode with zero momentum.
Therefore, we propagate the high energy solution of the Gross--Pitaevskii 
equation and, at each time step, do the decomposition into plane waves by 
using the Fourier transform technique in spatial domain. In such a way we gain 
the time dependence of natural modes. Their spectrum (Fourier transform with
respect to time) is plotted in Figs. \ref{spec1} and \ref{spec2} in the case of 
low and high temperatures, respectively, for modes $(0,0,1)$ and $(3,3,3)$. 
Typically, two groups of frequencies are visible in the spectrum and the 
separation between them increases with the mode energy. Moreover, increasing 
the energy of the mode leads also to the suppression of the lower frequencies 
group. Another important property is that the width of each group is getting 
broader when the temperature of the system is increasing.

The mean frequency of each group of frequencies is well described by the
gapless Bogoliubov--like formula: 
\begin{equation}
\epsilon_p = \sqrt{\left(\frac{p^2}{2}+gNn_0\right)^2-
\left(gNn_0\right)^2}  \;\;,
\label{Bogoliubov}
\end{equation}
where $n_0$ is the condensate fraction. This simple generalization of
the Bogoliubov formula is known as the Popov approximation
\cite{Pethick}. In Fig. \ref{exc}, which is the
case of $(1,1,1)$ mode, frequencies $\mu \pm \epsilon_p$ are marked by the 
vertical dotted lines whereas those given by the simplified expression 
(\ref{spectrum}) (when the driving force term is neglected) derived in the 
previous section by the dashed ones. Obviously, the formula (\ref{Bogoliubov}) 
works better. The excellent agreement between the Bogoliubov--like expression 
(\ref{Bogoliubov}) and numerical spectrum manifests itself in Fig. \ref{disp}, 
where we plot the quasiparticle energies for various temperatures (condensate 
populations) and interaction strengths. Numerical points (marked by boxes, circles, 
and triangles) are calculated as the mean frequency values (with respect to $\mu$) 
corresponding to higher frequency groups, whereas the solid lines come from 
(\ref{Bogoliubov}).

\begin{figure}[hbt]
\resizebox{2.8in}{2.4in}
{\includegraphics{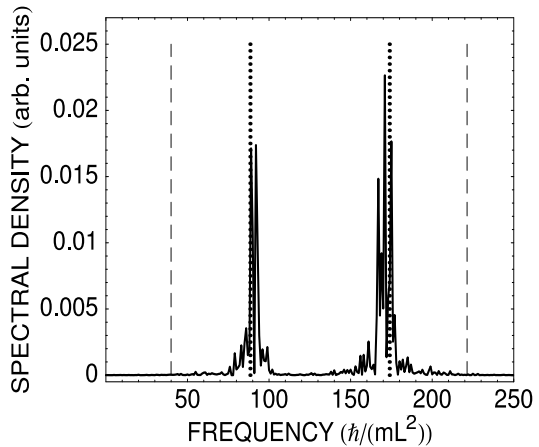}}
\caption{Fourier transform (in arbitrary units) of the (1,1,1) mode for
gN=731.5 and $n_0=0.79$. Dotted vertical lines show frequencies obtained
based on the formula (\ref{Bogoliubov}) whereas dashed ones mark frequencies 
calculated from the simplified expression (\ref{spectrum}). }
\label{exc}
\end{figure}

\begin{figure}[hbt]
\resizebox{3.8in}{2.5in}
{\includegraphics{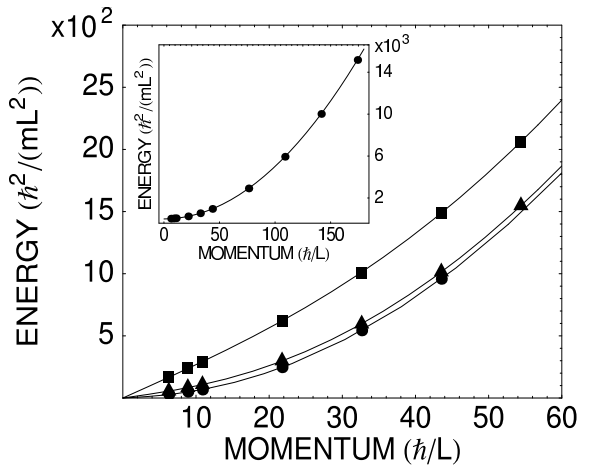}}
\caption{Dispersion relation for various interaction strengths and temperatures.
Numerical results are marked by boxes ($gN=731.5$, $n_0=0.95$), 
circles ($gN=73.15$, $n_0=0.13$), and triangles ($gN=73.15$, $n_0=0.93$).
Solid lines are obtained from the expression (\ref{Bogoliubov}).
Inset shows excellent agreement between the numerical data and formula
(\ref{Bogoliubov}) for all modes in the case of high temperature.}
\label{disp}
\end{figure}

\begin{figure}[thb]
\resizebox{2.8in}{2.4in}
{\includegraphics{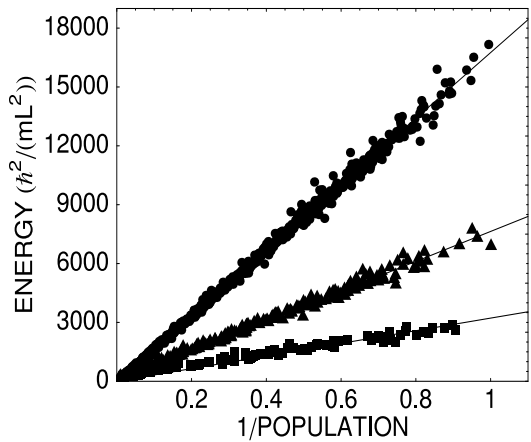}}
\caption{Quasiparticle energies as a function of inverse of the population
for the system of $N=235000$ ($gN=731.5$) atoms for various temperatures.
The upper set of data (circles) was obtained on the grid 36x36x36 and
corresponds to the temperature of $9.37$ nK ($n_0=0.07$), the middle one 
(triangles) is the case of $4.27$ nK ($n_0=0.79$) and 24x24x24 grid whereas 
the lower set (marked by boxes) results from the 16x16x16 grid and gives the 
temperature of $1.80$ nK ($n_0=0.95$). }
\label{tem}
\end{figure}

Approximate analytical solutions show that amplitudes 
$\alpha^{}_{\bf p}$ and $\alpha^{*}_{\bf -p}$ oscillate with two frequencies.
By taking appropriate linear combination of these amplitudes one
can find normal modes of the system, i.e., modes oscillating with only one
frequency. This transformation is just the transformation to Bogoliubov
quasiparticle amplitudes. They can be regarded as an elementary
excitations of the system. However, numerical results show that 
instead of two frequencies one obtains two distinct groups of sharp frequencies.
For each positive frequency component $\mu+\epsilon$ (where $\epsilon$
corresponds to the frequency within the peak) there exists companion at the frequency
$\mu-\epsilon$. Moreover, the phase difference of these components is equal to the
corresponding one for mode ${\bf -p}$. The center of the higher frequencies group
defines the energy of quasiparticle while the width of this group is related
to its lifetime. By integrating over the quasiparticle spectrum one can get the
occupation of the quasiparticle mode. In the phonon range the occupation
of the quasiparticle must be obtained with the help of the Bogoliubov
transformation. This task is simplified because of the phase relation mentioned
above. Having energies of elementary excitations and their populations
we can assign the temperature to the system \cite{Oxford2}. 
Since the modes are macroscopically occupied we expect that the
equipartition relation is fulfilled, i.e., $N n_{\bf p} (\epsilon_{\bf p} - \mu) =
k_B T$ for each mode. As it is the case shows Fig. \ref{tem} where we plotted the
energy of quasiparticles as a function of inverted population. Due to the
equipartition relation this dependence should be linear and the slope of the line
is just the temperature of the system. However, since the range of the 'y' axis 
in Fig. \ref{tem} is rather large some details for small momenta might be hidden.
Indeed, the presentation of the same results in a different form (see Fig.
\ref{tem1}) shows departure of low energy modes (phonons) from the equipartition
law. Finally, in
Fig. \ref{occup} we plot the condensate occupation for various temperatures for
the system built of $N=235000$ atoms. The dashed line in Fig. \ref{occup} shows
the population of an ideal condensate of the same density. Although we can not
precisely determine the value of critical temperature of finite size interacting
gas, Fig. \ref{occup} clearly shows that the shift of the critical temperature is
positive.

\begin{figure}[hbt]
\resizebox{2.8in}{2.4in}
{\includegraphics{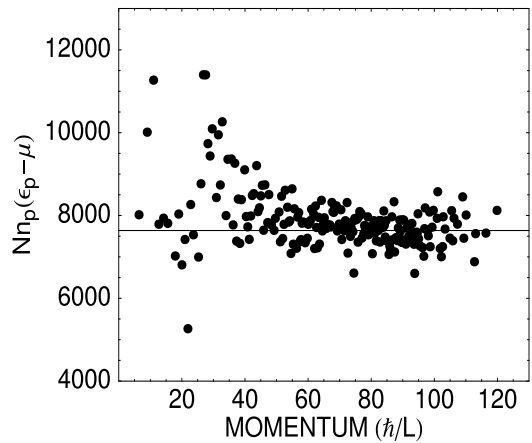}}
\caption{Verifying the equipartition relation we plot the expression
$N n_{\bf p} (\epsilon_{\bf p} - \mu)$ (in units of $\hbar^2/(mL^2)$)
for all modes. Note the imperfect equipartition for small momenta (phonon
modes). Parameters are the same as in the case of middle set of data in 
Fig. \ref{tem}.}
\label{tem1}
\end{figure}

\begin{figure}[hbt]
\resizebox{2.8in}{2.4in}
{\includegraphics{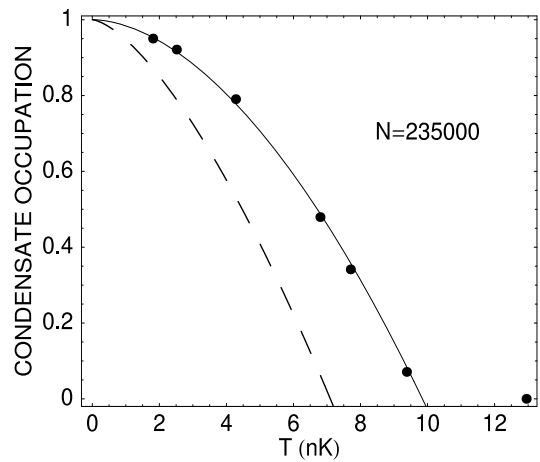}}
\caption{Condensate occupation as a function of temperature for a uniform
weakly interacting Bose gas; the scattering length equals $a_s=25$ nm
($gN=731.5$, $N=235000$). The dashed line shows the fraction of atoms in
the condensate for the ideal gas of the same density (the critical temperature 
is about $T_0 = 7.1$ nK. The solid line is the fit: $1-(T/9.85)^{1.80}$. }
\label{occup}
\end{figure}

\begin{figure}[hbt]
\resizebox{3.8in}{2.5in}
{\includegraphics{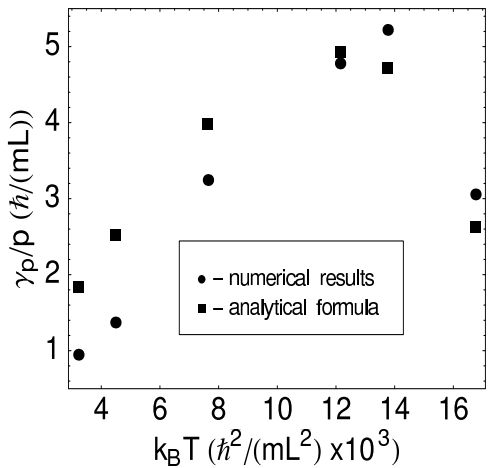}}
\caption{Comparison of the ratio $\gamma_p/p$ calculated numerically
(circles) with that coming from the analytical formula (\ref{damping})
(and given by boxes) for various temperatures and $gN=731.5$.}
\label{damp2}
\end{figure}

\begin{figure}[hbt]
\resizebox{2.8in}{2.4in}
{\includegraphics{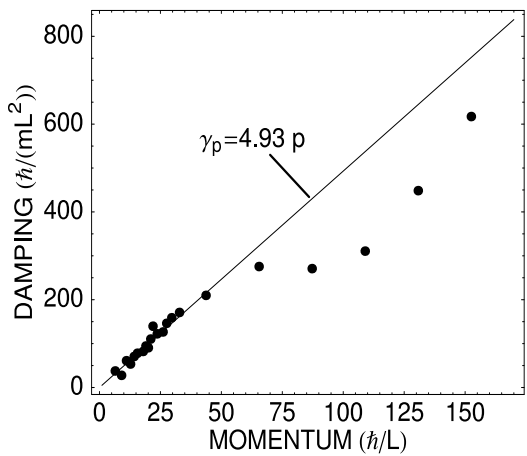}}
\caption{The decay rates as a function of momentum for $gN=731.5$
and higher temperature ($n_0=0.48$).
Both phonon--like and particle--like parts of spectrum are considered. 
Solid line follows the analytical formula (\ref{damping}). }
\label{damp}
\end{figure}

Finally, we would like to discuss the damping of the quasiparticle modes.
There exists analytical result for the decay rates of elementary excitations
in a uniform Bose gas. At high temperatures ($k_B T \gg gN/V$) Landau processes 
dominate and the decay rate for the phonon--like part of the spectrum follows 
the formula \cite{Landau}:
\begin{equation}
\gamma_p = \frac{3\pi^{3/2}}{2} \frac{k_B T}{gN/V} 
\left(\frac{N}{V}\, a^3_s\right)^{1/2} \frac{\epsilon_p}{\hbar} \;.
\label{damping}
\end{equation}
Since phonon energies depend linearly on the momentum, so do the decay rates.
The ratio $\gamma_p/p$ is a function of temperature given by 
$\sim\!\! \sqrt{n_0}\, T$,
where $n_0$ is the condensate fraction. In Fig. (\ref{damp2}) we compare 
numerical results with the analytical expression (\ref{damping}). In order to
determine the damping rates we average the spectrum of phonon--like modes over 
the angles in the momentum space and fit to the Lorentzian curve. The FWHM of 
the Lorentzian fit is just the damping rate. We see that both approaches agree 
quite well except of two first points. 
These points correspond to very low temperatures (approximately $2000$ and
$4000$ in units of $\hbar^2/(mL^2)$) and certainly the condition of the
validity of analytical formula (\ref{damping}) is not fulfilled in this case since
$gN/V$ is as high as $731.5$.
However, in our method we can go beyond the phonon--like excitations and find 
the decay rates for particle--like part of the spectrum. In Fig. \ref{damp} we 
plot the damping rates for all momenta in the case of higher temperature. The 
solid line follows the analytical expression (\ref{damping}). The numerical 
results and formula (\ref{damping}) agree only for small momenta as it should be 
expected (see Fig. \ref{damp2}). The damping rates of particle--like 
excitations show the nonlinear dependence on the momentum.

\section{Conclusions}
\label{concl}
We have applied the classical fields approach to the weakly interacting Bose gas
in a box with periodic boundary conditions. By Fourier transform we analyze the
frequency dependence of amplitudes of eigenmodes of the system. We observe that
the zero--momentum component of the field oscillates with single frequency which
is identified as a chemical potential. We have obtained nonperturbative results
for the chemical potential. For the physically relevant parameters it differs
strongly from the simple two--gas formula. All other modes have much more 
complicated time evolution. Their spectra consist of two groups of frequencies
displaced symmetrically with respect to the chemical potential. The center of
gravity of each groups of frequencies agrees perfectly with a simple formula
known as the Popov formula \cite{Pethick}. The low momenta excitations have a 
phonon--like dispersion while for high momenta components the dispersion is of 
particle kind. Knowing the excitation spectrum of quasiparticles we are able to 
find also their populations and we can test the equipartition of the energy 
between the modes and then determine the temperature. The width of the spectrum 
is identified as the damping rate of quasiparticles. We observe that our 
nonperturbative results agree reasonably well with the analytic estimates for 
the phonon damping rates for high temperatures.

\acknowledgments
M.B. and M.G. acknowledge support by the Polish KBN Grant No. 2 P03B 052 24.
K.R. is supported by the subsidy of the Foundation for Polish Science.
Some of our results have been obtained using computers at the Interdisciplinary 
Centre for Mathematical and Computational Modelling of Warsaw University.
This research was partly supported by the Polish Ministry of Scientific 
Research Grant Quantum Information and Quantum Engineering.


\begin{thebibliography}{99}

\bibitem{BEC} M.H. Anderson, J.R. Ensher, M.R. Matthews, C.E. Wieman,
and E.A. Cornell, Science {\bf 269}, 198 (1995);
K.B. Davis, M.-O. Mewes, M.R. Andrews, N.J. van Druten, D.S. Durfee,
D.M. Kurn, and W. Ketterle, Phys. Rev. Lett. {\bf 75}, 3969 (1995);
C.C. Bradley, C.A. Sackett, J.J. Tollett, and R.G. Hulet, 
Phys. Rev. Lett. {\bf 75}, 1687 (1995) and Erratum {\bf 79}, 1170(E)
(1997).

\bibitem{Stringari} F. Dalfovo, S. Giorgini, L.P. Pitaevskii, and S. Stringari,
Rev. Mod. Phys. {\bf 71}, 463 (1999).

\bibitem{Zaremba} 
E. Zaremba, A. Griffin, and T. Nikuni, Phys. Rev. A {\bf 57}, 4695 (1998);
B. Jackson and E. Zaremba, Phys. Rev. Lett. {\bf 88}, 180402 (2002);
B. Jackson and E. Zaremba, Phys. Rev. A {\bf 66}, 033606 (2002);
B. Jackson and E. Zaremba, New J. Phys. {\bf 5}, 88 (2003).

\bibitem{classf} B.V. Svistunov, J. Moscow Phys. Soc. {\bf 1}, 373 (1991);
K. Damle, S.N. Majumdar, and S. Sachdev, Phys. Rev. A {\bf 54}, 5037 (1996);
Yu. Kagan and B.V. Svistunov, Phys. Rev. Lett. {\bf 79}, 3331 (1997);
N.G. Berloff and B.V. Svistunov, Phys. Rev. A {\bf 66}, 013603 (2002).

\bibitem{OpEx} K. G\'oral, M. Gajda, and K. Rz\c a\.zewski,
Opt. Express {\bf 8}, 92 (2001).

\bibitem{Oxford1} M.J. Davis, S.A. Morgan, and K. Burnett, Phys. Rev. Lett.
{\bf 87}, 160402 (2001); 
M.J. Davis, R.J. Ballagh, and K. Burnett, J. Phys. B {\bf 34}, 4487 (2001).

\bibitem{Oxford2} M.J. Davis, S.A. Morgan, and K. Burnett, Phys. Rev. A {\bf 66}, 
053618 (2002).

\bibitem{Davis} M.J. Davis and S.A. Morgan, cond-mat/0307155.

\bibitem{RapCom} K. G\'oral, M. Gajda, and K. Rz\c a\.zewski,
Phys. Rev. A {\bf 66}, 051602 (2002).

\bibitem{Castin}
I. Carusotto and Y. Castin, J. Phys. B {\bf 34}, 4589 (2001);
A. Sinatra, C. Lobo, and Y. Castin, J. Phys. B {\bf 35}, 3599 (2002);
I. Carusotto and Y. Castin, Phys. Rev. Lett. {\bf 90}, 030401 (2003).

\bibitem{Pethick} see for instance C. J. Pethick and H. Smith, 
{\it Bose--Eintein Condensation in Dilute Gases} 
(Cambridge University Press, New York, 2002), p. 225. 

\bibitem{Schmidt} H. Schmidt, K. G\'oral, F. Floegel, M. Gajda, 
and K. Rz\c a\.zewski, J. Opt. B {\bf 5}, S96 (2003).

\bibitem{gapless} N.M. Hugenholtz and D. Pines, Phys. Rev. {\bf 116},
489 (1959).

\bibitem{Landau} 
P.O. Fedichev, G.V. Shlyapnikov, and J.T.M. Walraven, Phys. Rev. Lett. {\bf 80},
2269 (1998);
P.O. Fedichev and G.V. Shlyapnikov, Phys. Rev. A {\bf 58},
3146 (1998).


\end{thebibliography}
\end{document}